**A New Way to Look at Regional Survey Data: Differences in Vacancy Rates and Persons per Household by County, 2000-2005**

Charles D. Coleman[a]* and Jonathan F. Takeuchi[b]

[a]*Economic Statistical Methodology Division, U.S. Census Bureau, Washington, DC, https://orcid.org/0000-0001-6940-8117, https://www.linkedin.com/in/chuckcoleman;* [b]*Willis Towers Watson, Tokyo, Japan, https://www.linkedin.com/in/jonathan-takeuchi-3401679*

* Economic Statistical Methodology Division, U.S. Census Bureau, Washington, DC, 20233, charles.d.coleman@census.gov

# A New Way to Look at Regional Survey Data: Differences in Vacancy Rates and Persons per Household by County, 2000-2005

Regional survey estimates and their significance levels are simultaneously displayed in maps that show all 3,141 U.S. counties and equivalents. An analyst can focus his attention on significant differences (or those with a different, low-valued uncertainty measure) for all but the very smallest counties. Differences between Census 2000 and the 2005 American Community Survey values are shown.

## 1 Introduction

Many surveys produce estimates and uncertainty measures for a large number of geographic units. We introduce a mapping technique that can simultaneously map both while providing a means for "uninteresting" estimates to not be displayed. This enables the analyst to focus his attention on areas with significantly different values, however defined, and to ignore extraneous, insignificant values, without zooming except for the very smallest areas. We map difference differences in persons per household (PPH) and vacancy rates between 2005 American Community Survey (ACS) estimates and Census 2000 for all 3,141 counties and county equivalents. We choose 2005 because it is the year the ACS was implemented nationwide, making data for all counties are available. It also was the latest year available to us at the time of writing. Nationwide, comparison of the two datasets shows PPH is lower in 2005 while the vacancy rate is higher in 2005. However, what is true nationally is not necessarily true for any particular subnational geographic area. We analyze differences in PPH and vacancy rates, that is, the 2005 ACS values less the Census 2000 values of these variables, at the county level for all counties, together with their uncertainties, as measured by *p*-values. (From here on, we simply refer to counties and county equivalents as "counties," using the singular when necessary.) While we use *p*-values, other uncertainty measures like Bayes factors or coefficients of variation can be used. We make no attempt to decompose the differences between true change from 2000 to 2005, coverage differences between Census 2000 and the ACS, and other nonsampling errors.

The ability to map large numbers of areas, large and small, is in contrast to the methods discussed in Francis et al. (2015) which can only examine a small number of relatively large areas. Thus, we overcome the constraint noted by Torrieri, Wong and Ratcliffe (2001, p. 1098) that maps conveying both level and uncertainty for all counties in the U.S. are incomprehensible to most viewers.

## 2 Data Sources

Data for 2000 come from Census 2000 100 percent (i.e., short form) data (U.S. Census Bureau 2006a). Data for 2005 are unpublished estimates based on the full sample used in the 2005 American Community Survey (U.S. Census Bureau 2006b). The ACS data were weighted using uncontrolled sample weights, adjusted for nonresponse. That is, the ACS data used are not controlled to any external housing unit or household population counts in order to have pure survey estimates. ACS values used here may vary from published data due to different weighting schemes.

## 3 Methodology

SAS procedure PROC SURVEYMEANS computed the 2005 survey estimates nationally and by county. The standard errors were computed using the replicate weights using the methodology described in U.S. Census Bureau (2006b, pp. 12-1—12-2). *P*-values were derived from the two using the standard normal distribution.

Census 2000 is not strictly compatible with the 2005 ACS due to differences in coverage rates of housing units and population and other nonsampling errors. Therefore, when there is no true change in some of the variables, some differences, many significant, are to be expected. The reader is cautioned not to interpret the differences in terms of true change or conceptual or other differences between Census 2000 and the ACS.

We separately map differences and their *p*-values. Then, the two variables are mapped simultaneously using a color coding. These last two maps accomplish the objective of displaying a variable with an associated uncertainty measure.

## 4 National Results

We begin by assessing differences at the national level using standard statistical techniques. Nationwide, the vacancy rate rose by 1.0 percentage points from 9.0 percent in Census 2000 to 10.9 percent in the 2005 ACS and PPH fell by 0.08 from 2.59 persons in Census 2000 to 2.51 persons in the 2005 ACS. (The first difference shown is not equal to the difference of the two vacancy rate values due to rounding.) Table 1 displays these values, along with statistical tests. Both differences are with *p*-values less than 0.01% in two-tailed tests. The sample sizes in the 2005 ACS are 1,902,714 surveyed housing units, of which 1,768,518 were occupied. The sample weights adjust for the variable probabilities of sampling and sum up to the total number of housing units eligible for surveying in the 2005 ACS for the vacancy rate data and to the estimated total number of households for PPH. Since what may be true at the national level is not necessarily true for subnational geographic areas, succeeding analyses use county-level data.

## 5 Interpreting Maps of County Differences

Figures 1 and 2 map estimated differences, the 2005 ACS values less the Census 2000 values, in vacancy rates and PPH by county. Large positive differences are shown in blue, small positive differences in green, small negative differences in orange, and large negative differences in red. These maps show considerable local variation in the estimated differences. In many cases, a county with a large, positive estimated difference, that is, its 2005 ACS PPH or vacancy rate is much greater than its Census 2000 value, is adjacent to a county with a large, negative estimated difference, that is, its 2005 ACS PPH or vacancy rate is much lower than its Census 2000 value. These show up on the maps as adjacent blue and red patches, respectively. This appears contrary to an assumption of smoothness: one county's difference generally should not be radically different from those of its neighbors and the existence of these radical differences should be few in number. However, the problem with attempting to interpret these maps is that the differences in most counties are not significant. We simply cannot tell which of these blue-red county pairs reflect statistically significant differences and which are due to chance only. Moreover, we cannot tell if the general trend of negative PPH differences, shown as red and orange in Figure 2, reflect statistically significant differences. Therefore, we cannot draw meaningful conclusions about county differences from these maps alone.

## 6 Locating Significant Differences

This section is concerned with geographically locating significant differences and determining whether or not those differences as a whole are due to chance alone. The second task is tackled first. The primary tool is the one-sided $p$-value, from which two-sided significance levels are derived. The one-sided $p$-value is the probability that, under the null hypothesis ($H_0$) of zero true difference, chance produces a value less than or equal to the observed difference. Because *p*-values are probabilities, they always lie between 0 and 1. The global null hypothesis ($G_0$) is that all of the differences for a variable are due to chance alone. If the observed *p*-values are indistinguishable from a uniform distribution, then $G_0$ is rejected.

$G_0$ is tested by examining tables and plots of the *p*-values and by explicit Sign Tests of the proportion of observations with two-sided significance values of 10% or less. Tables 2 and 3 show summary statistics for the *p*-values for vacancy rates and PPH, respectively. The *p*-values are grouped into ranges corresponding to the conventional two-sided maximum significance levels 1%, 5%, and 10%. The expected percentages under $G_0$ are the expected percentages of *p*-values falling into each range under $G_0$, that is, if they were generated by chance alone. Observing greater than expected amounts of *p*-values in the significant tail ranges is an indication that the observed differences are not due to chance alone. Vacancy rate differences have greater than expected *p*-value counts in each of the ranges greater than .95 and in the one range less than .005. PPH differences show greater than expected *p*-value counts in all of the *p*-value ranges less than .05. This strongly suggests that we can reject $G_0$ for both vacancy rate and PPH differences.

Tables 4 and 5 display the counts of maximum conventional two-sided significance levels, obtained, for each level for vacancy rate and PPH differences, respectively, by summing the counts of *p*-values for the corresponding two *p*-value ranges. Even including the tails that lack significant observations does not alter the finding of excessive counts of significant observations relative to the null hypothesis. For example, 9.51 percent and 12.77 percent of nonzero vacancy rate and PPH differences are significant at the 1% level.

Figures 3 and 4 display QQ-plots of the *p*-values for vacancy rate and PPH differences, respectively. The red lines illustrate the uniform distribution of the *p*-values under $G_0$. The observations (i.e., counties) are plotted as plus signs, giving the appearance of an irregular, black line. The observations clearly deviate from the red line. Therefore, the finding of large numbers of significant differences is not due to chance alone and we reject the global null hypothesis $G_0$. Further confirmation can be seen in Table 6, which shows *z*-scores of the Sign Tests for $G_0$ for both vacancy rate and PPH differences. From the standard normal distribution, their *p*-values are both well below .0001, even when using the correct one-sided interpretation of the Sign Test. Thus, $G_0$ is rejected for both variables.

We can make further inferences from these Figures. Figure 3 shows observations lying generally above the red line, indicating a general trend of increasing vacancy rates. However, the very lowest set of observations lies below the red line, indicating that a smaller number of significant decreases are also present. Figure 4 shows observations lying below the red line, indicating an overall decreasing trend in PPH. The few significant positive differences ($p > .95$) may be spurious, arising only from chance.

Finally, Figures 5 and 6 map the *p*-values for vacancy rate and PPH differences, the 2005 ACS estimates less the Census 2000 values, respectively. A county with a significant *p*-value (using a two-sided test) has a shade of red, if the difference is negative, or a shade of blue, if the difference is positive. These can be used together with the maps in Figures 1 and 2 to identify significant variables and their levels. However, this is a difficult task, as the analyst has to compare two maps, keeping in mind a large amount of data: in this case, 3000+ differences and their *p*-values. To make the analysis simpler, the combination of a variable and its significance level are mapped and discussed in the next section.

## 7 Maps Combining Survey Estimates and their Significance Levels

Our maps code estimates by hue and two-sided significance by saturation. The hues are identical to those used in Figures 1 and 2. The lower the significance level that an estimate has for an area, the more saturated is the color displayed for that area. Values that are not significant are left unshaded, showing that they are statistically indistinguishable from zero. This method of mapping data can be used for any set of regional survey estimates. While this method has previously been used with two variables (Brewer & Suchan, 2001: 12), to the authors' knowledge, it has not been used to display statistical uncertainty. The biggest gain from using this approach is to ignore highly uncertain estimates, thereby focusing the maps on values with least uncertainty, measured by *p*-values in these examples.

Figures 7 and 8 simultaneously map vacancy rate and PPH differences and their significance levels. These maps, when compared to Figures 1 and 2, show many interesting things. Figure 1 shows that Sweetwater County, WY, the red-shaded county in southwestern Wyoming, has a large estimated vacancy rate decline. Two of its adjoining counties in Wyoming and its two southern neighbors, one each in Utah and Colorado, have large estimated vacancy rate increases, as shown by their blue color. Moreover, the rest of Wyoming is a patchwork of varying estimated vacancy rate differences. Figure 7 shows that most of these vacancy rate differences are statistically indistinguishable from chance variations. Sweetwater is the only county in Wyoming with a significant vacancy rate difference. Neither of its southern neighbors has a significant difference, as can be seen in Figure 9, a zoomed-in version of Figure 7.

Thus, the original finding that the direction of most of Sweetwater's neighboring counties' differences is opposite that of Sweetwater's difference is statistically rejected.

An opposite example is Summit, UT's PPH, which has a positive difference, while many surrounding counties have negative differences. In this case, this positive difference and many of the surrounding negative differences are statistically significant, as shown in Figure 10, an enlarged view of Figure 8.

These maps demonstrate that having a large population and, therefore, a large sample size, is no guarantee of significance. For example, Anchorage Borough, AK in south-central Alaska has an estimated population of about 275,000 people in 2005 (U.S. Census Bureau, 2006b), but its estimated vacancy rate difference is not significant. On the other hand, Denali Borough, in central Alaska with an estimated population of less than 2,000 people (U.S. Census Bureau, 2006b), has a large vacancy rate difference, significant at 1%. Anchorage's vacancy rate difference is small: less than 0.05 percentage point, causing it to be insignificant. These counties can be seen in Figure 11.

## 8 Conclusions

We have developed maps of regional survey data that enable an analyst to view both a variable and a measure of its uncertainty for large numbers of small geographic areas. They let the analyst focus on analyzing data with the least uncertainty, here measured by *p*-values, while not displaying data with too much uncertainty. This method is not without its challenges. One of these is to find an effective color scheme that allows each hue to represent a different variable range and each saturation level to represent a different significance level. The uncertainty variable need not be *p*-values, but can be coefficients of variation, Bayes' factors or posterior probabilities, to give a few examples.

This particular application has shown that ACS 2005 vacancy rate and PPH estimates are significantly different from Census 2000 values in far more counties than would be anticipated by chance alone, about 23 percent and 30 percent of county vacancy rate and PPH differences, respectively, are significant at 10% or less. Each significant difference represents a sum of true change, conceptual differences between Census 2000 and the ACS, and random sampling error. Further analysis using the Census 2000 Supplementary Survey (C2SS) could decompose the significant differences between estimates of true change and differences between Census 2000 and C2SS for those counties that were included in C2SS.

We showed zoomed maps for small areas. We, therefore, recommend using an interactive maps whenever there are geographic areas too small to be displayed on a single map. This also has presentational uses in general.

## Disclaimer

Any views expressed are those of the authors and not necessarily those of the Census Bureau. The Census Bureau has reviewed this product for unauthorized disclosure of confidential information. (Approval ID: CBDRB-FY21-ESMD-BXXXX.)

Table 1. National Vacancy Rate and PPH Statistics.

| | Vacancy Rate | PPH |
|---|---|---|
| ACS 2005 Estimate | 0.109 | 2.511 |
| Standard Error | <0.001 | 0.001 |
| Census 2000 Value | 0.090 | 2.594 |
| Difference, 2000 to 2005 | 0.010 | -0.083 |
| *t* (Difference) | 277.12 | –56.09 |
| *p* (two-sided) | <.0001 | <.0001 |
| *N* | 1902714 | 1768518 |
| Sum of weights | 122592161 | 109191370 |

Sources: U.S. Census Bureau, American Community Survey 2005, Census 2000

Table 2. *p*-values of County Vacancy Rate Differences.

| *p*-value | Count | Percent | Expected Percent under $G_0$ |
|---|---|---|---|
| 0.995+ | 220 | 7.02 | 0.5 |
| 0.975–0.995 | 201 | 6.41 | 2.0 |
| 0.95–0.975 | 162 | 5.17 | 2.5 |
| Other | 2370 | 75.62 | 90.0 |
| 0.025–0.05 | 54 | 1.72 | 2.5 |
| 0.005–0.025 | 49 | 1.56 | 2.0 |
| <0.005 | 78 | 2.49 | 0.5 |

Sources: U.S. Census Bureau, American Community Survey 2005, Census 2000

Note: *p*-values for only 3,134 of 3,141 counties are available. Seven counties have estimated 2005 vacancy rates of zero, making their estimated standard errors zero, prohibiting calculation of *p*-values. All *p*-values are one-sided.

Table 3. *p*-values of County PPH Differences.

| *p*-value | Count | Percent | Expected Percent under $G_0$ |
|---|---|---|---|
| 0.995+ | 0 | 0 | 0.5 |
| 0.975–0.995 | 11 | 0.35 | 2.0 |
| 0.95–0.975 | 15 | 0.48 | 2.5 |
| Other | 2162 | 68.83 | 90.0 |
| 0.025–0.05 | 230 | 7.32 | 2.5 |
| 0.005–0.025 | 322 | 10.25 | 2.0 |
| <0.005 | 401 | 12.77 | 0.5 |

Sources: U.S. Census Bureau, American Community Survey 2005, Census 2000

Note: All *p*-values are one-sided.

Table 4. Two-Sided Significance Levels for County Vacancy Rate Differences.

| Significance Level | Count | Percent |
|---|---|---|
| 1% | 298 | 9.51 |
| 5% | 250 | 7.98 |
| 10% | 216 | 6.89 |
| Not significant | 2370 | 75.62 |

Sources: U.S. Census Bureau, American Community Survey 2005, Census 2000

Note: Significance levels shown are the minimum conventional significance levels of the data.

Table 5. Two-Sided Significance Levels for County PPH Differences

| Significance Level | Count | Percent |
|---|---|---|
| 1% | 401 | 12.77 |
| 5% | 333 | 10.60 |
| 10% | 245 | 7.80 |
| Not significant | 2162 | 68.83 |

Sources: U.S. Census Bureau, American Community Survey 2005, Census 2000

Note: Significance levels shown are the minimum conventional significance levels of the data.

Table 6. Sign Test Values for $G_0$.

| Variable | *z* |
|---|---|
| Vacancy Rates | 26.83 |
| PPH | 39.55 |

Note: Due to the large sample sizes ($n > 3,000$), the normal approximation is used.

Figure 1. Differences in Vacancy Rates: 2005 American Community Survey Less Census 2000

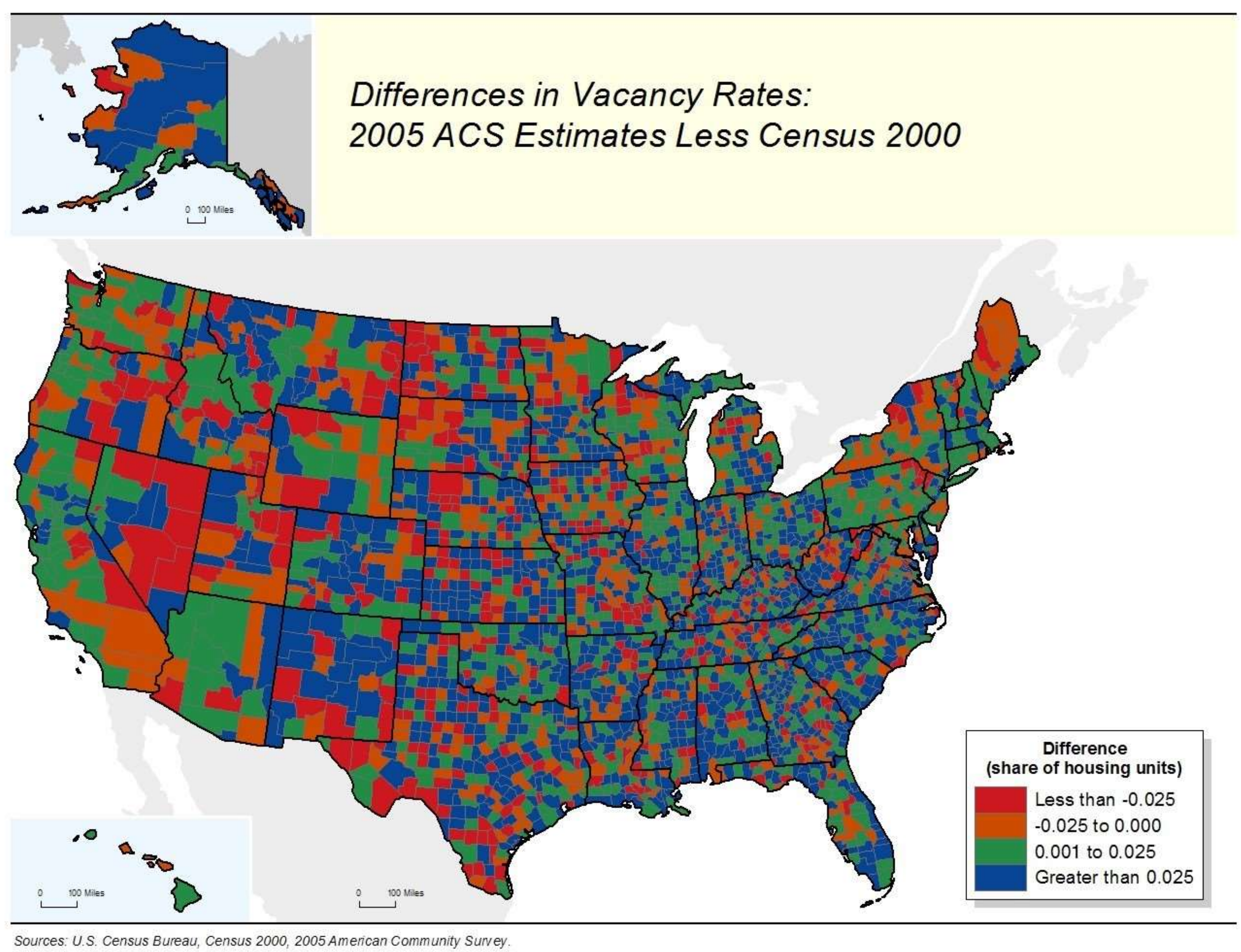

Figure 2. Differences in Persons Per Household: 2005 American Community Survey Less Census 2000

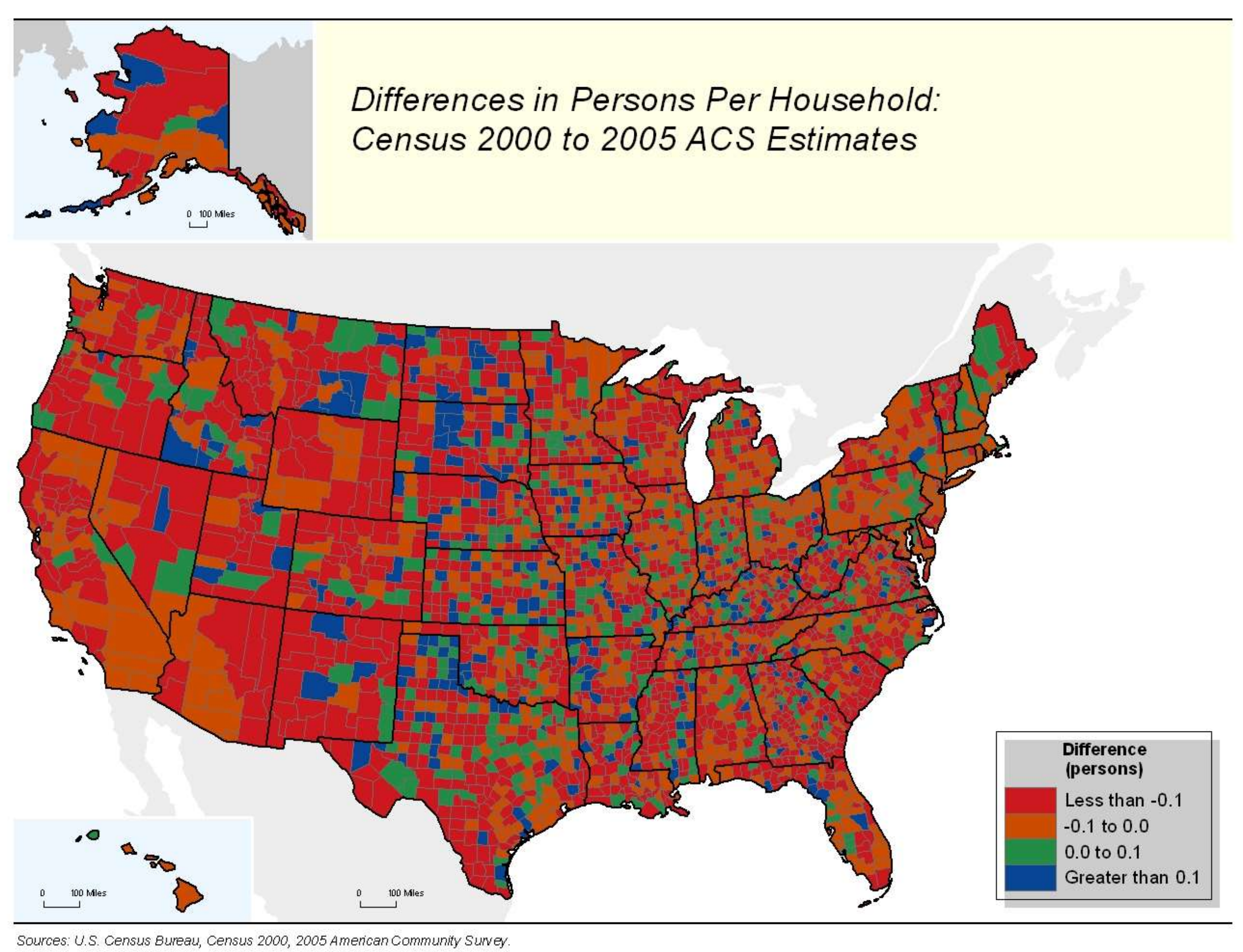

Figure 3. Q-Q Plot of County Vacancy Rate *p*-Values.

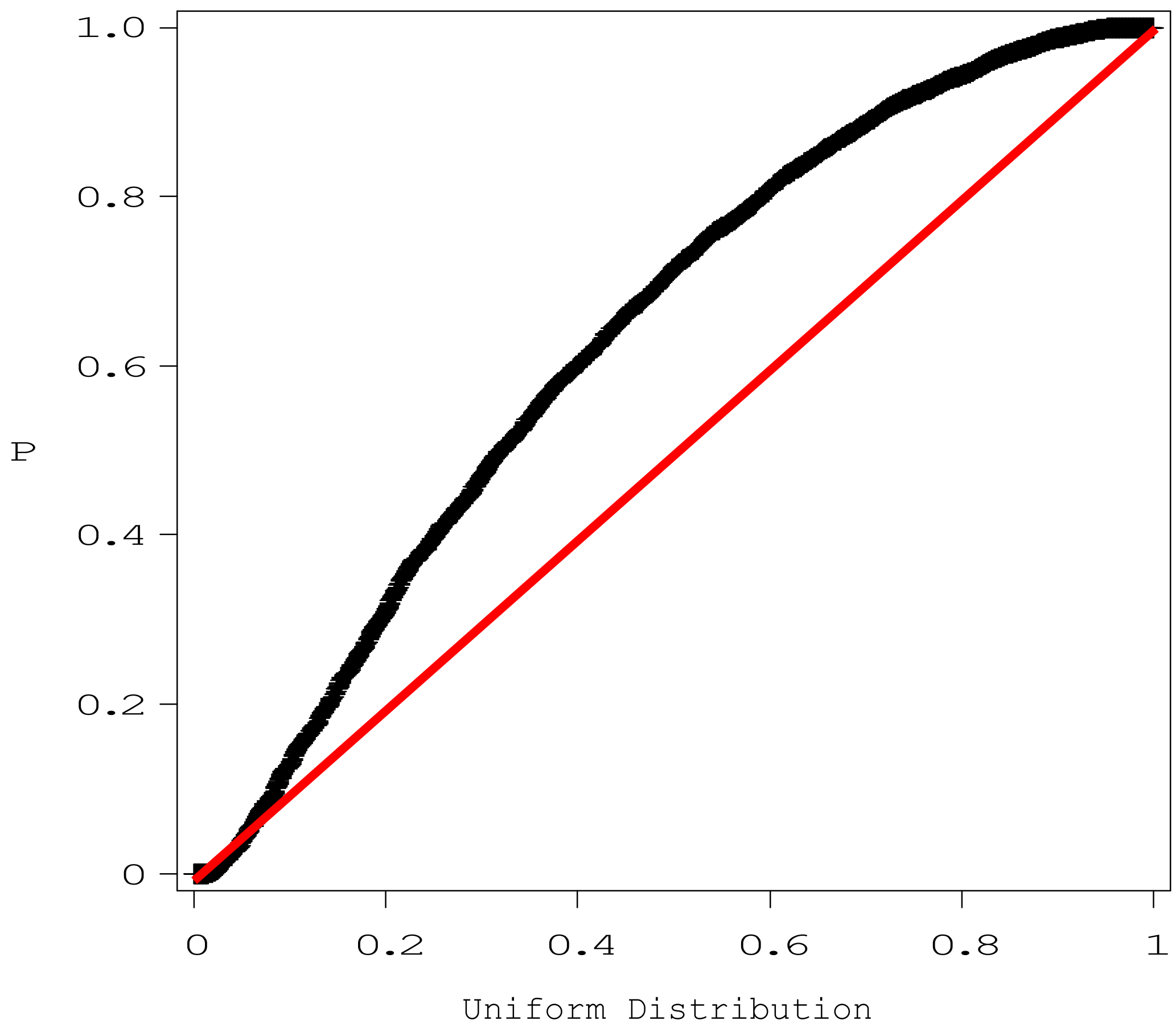

Figure 4. Q-Q Plot of County PPH $p$-Values.

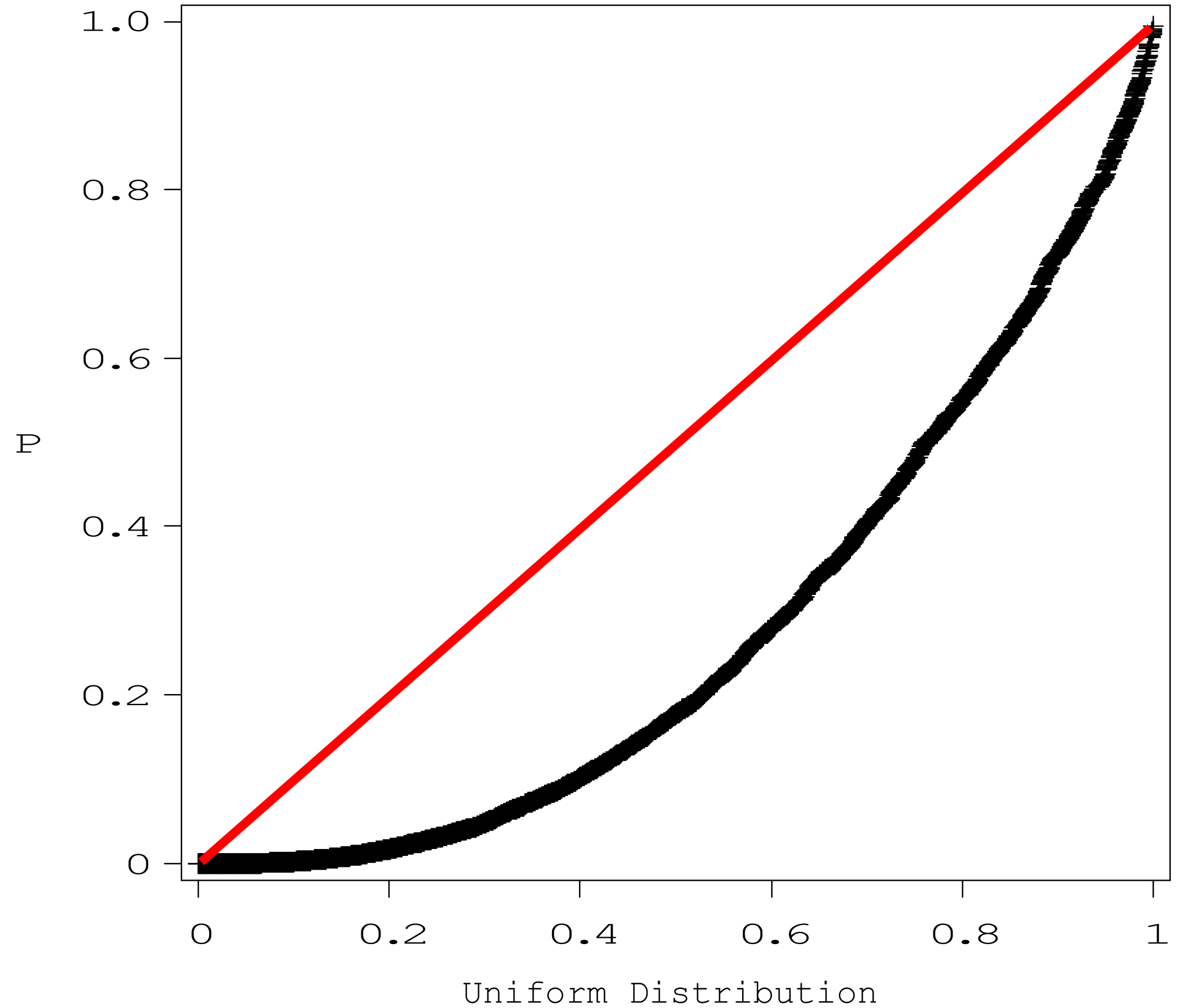

Figure 5. $p$-Values of Differences in Vacancy Rates: 2005 ACS Estimates Less Census 2000.

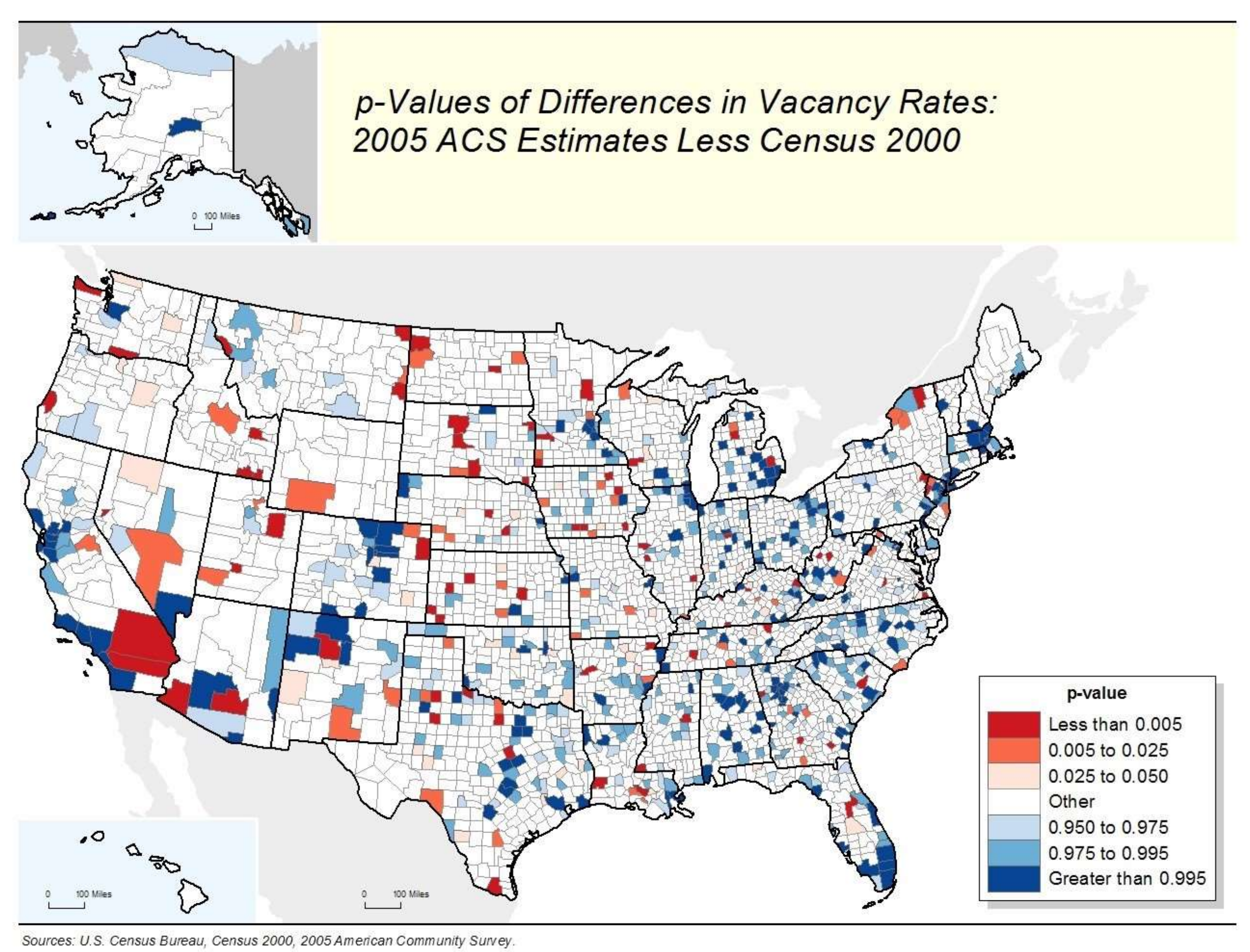

Figure 6. $p$-Values of Differences in Persons Per Household: 2005 ACS Estimates Less Census 2000.

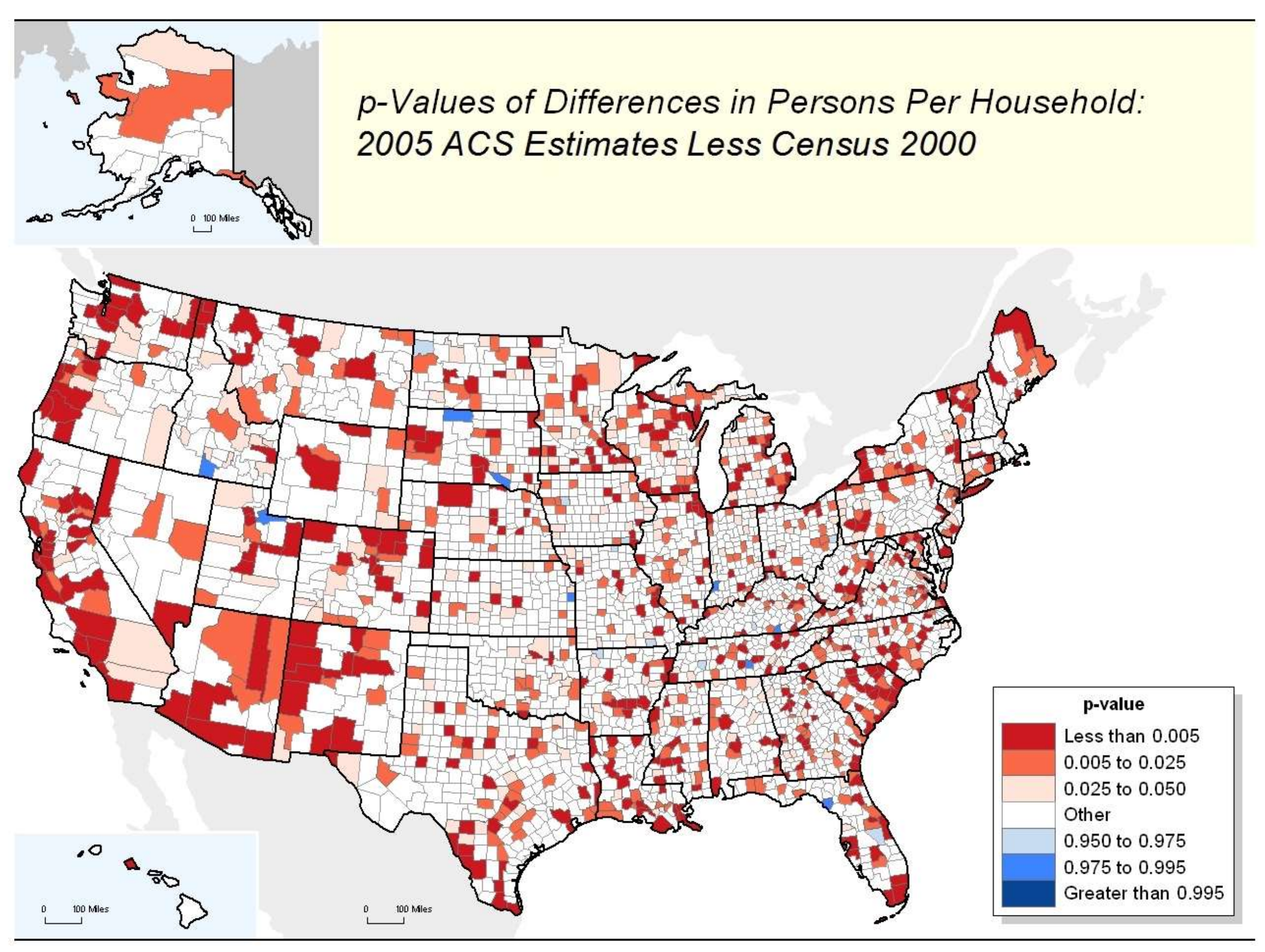

Figure 7. Significant Differences in Vacancy Rates: 2005 ACS Estimates Less Census 2000.

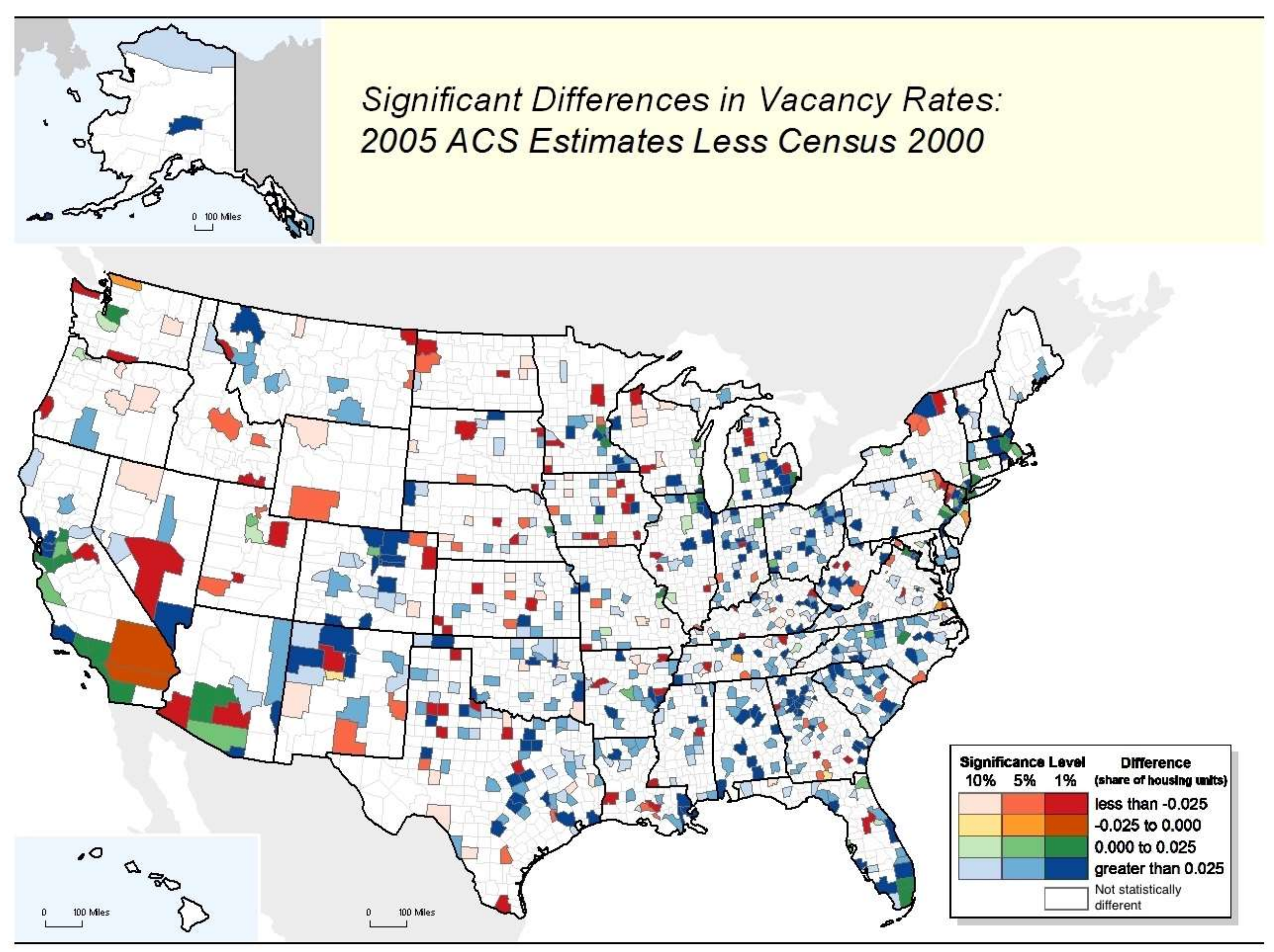

Figure 8. Significant Differences in Persons Per Household: 2005 ACS Estimates Less Census 2000.

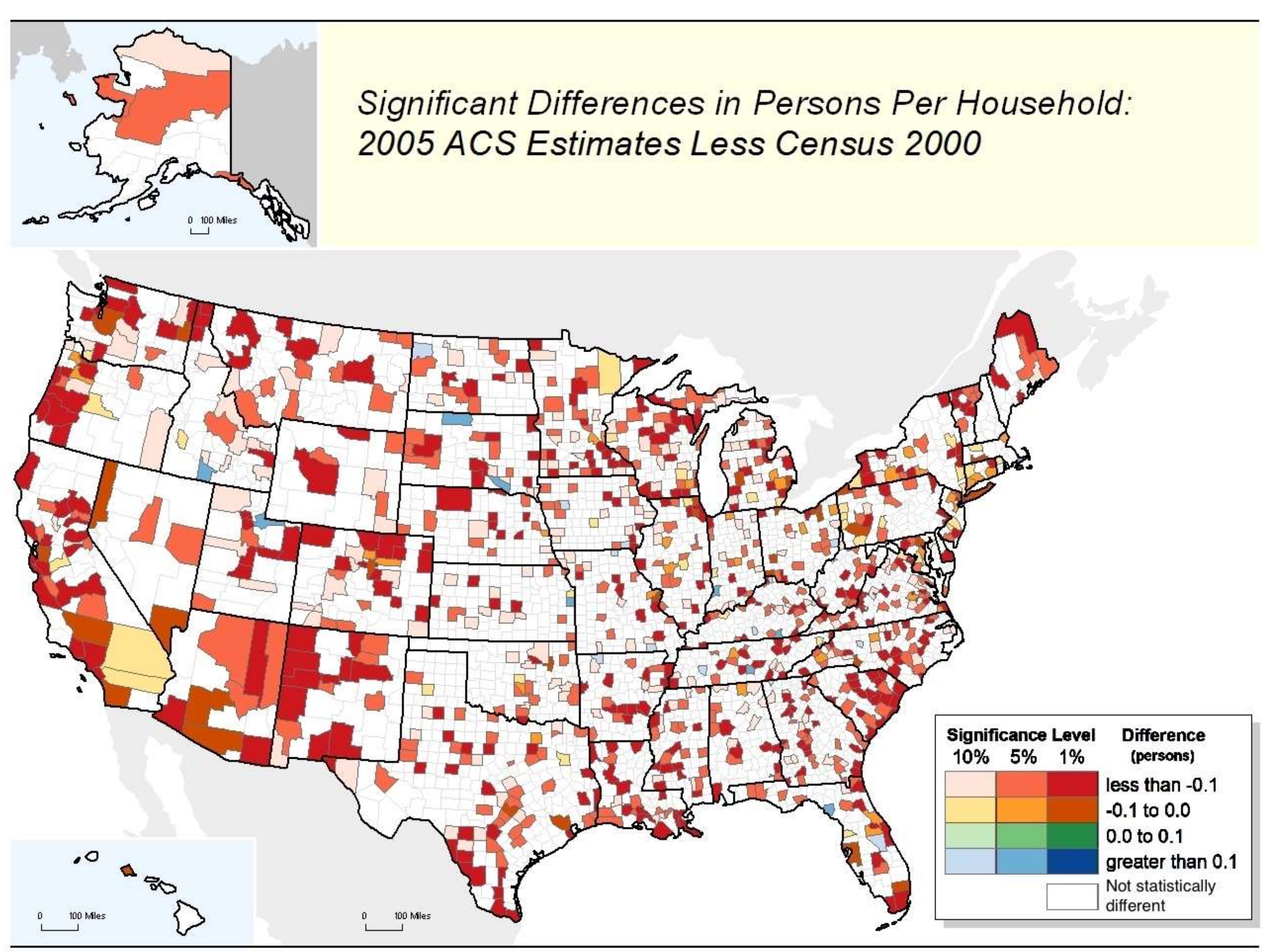

Figure 9. Significant Differences in Vacancy Rates: 2005 ACS Estimates Less Census 2000.

***Significant Differences in Vacancy Rates:***
***2005 ACS Estimates Less Census 2000***

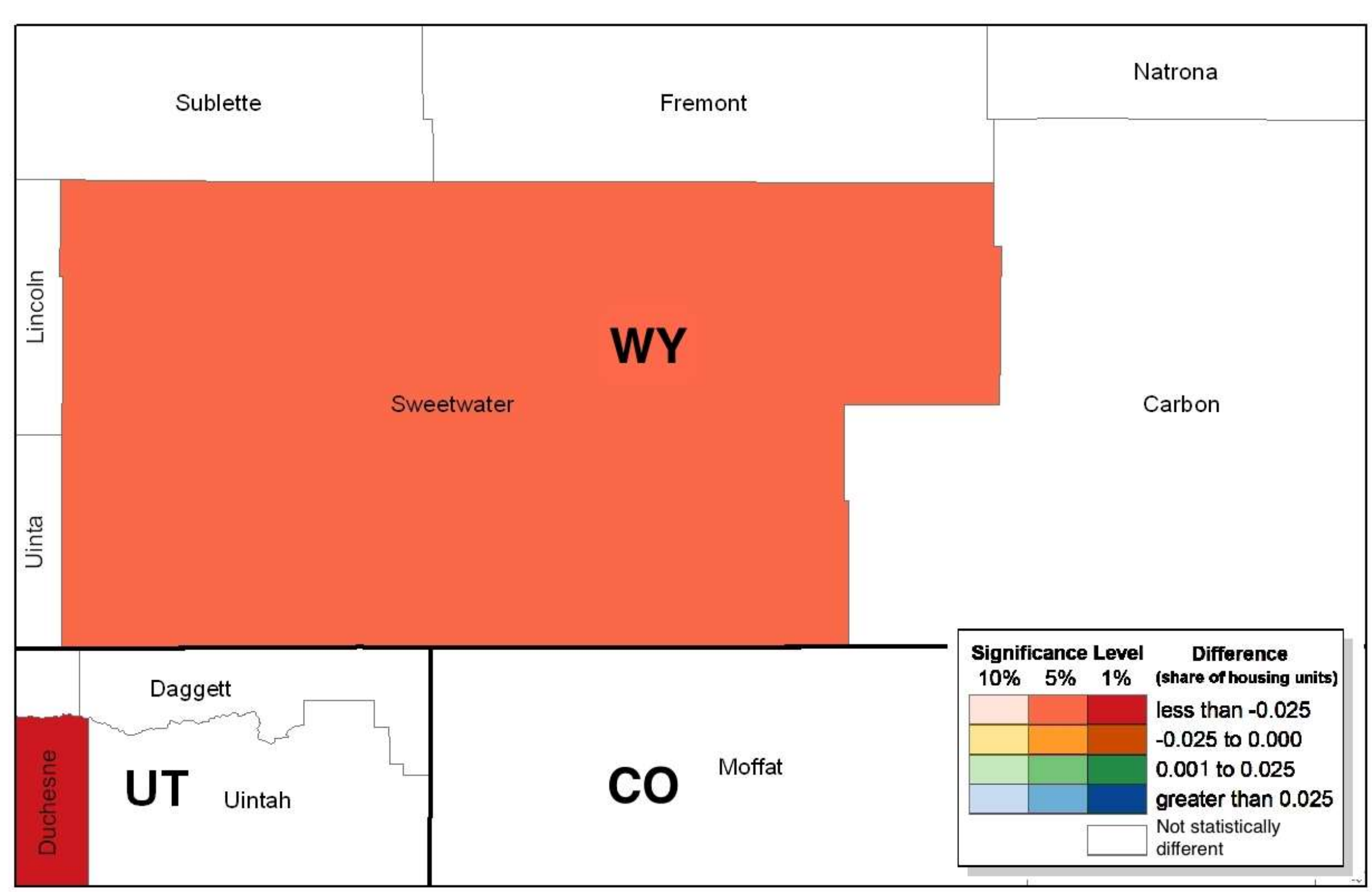

Figure 10. Significant Differences in Persons Per Household: 2005 ACS Estimates Less Census 2000.

***Significant Differences in Persons Per Household: Census 2000 to 2005 ACS Estimates***

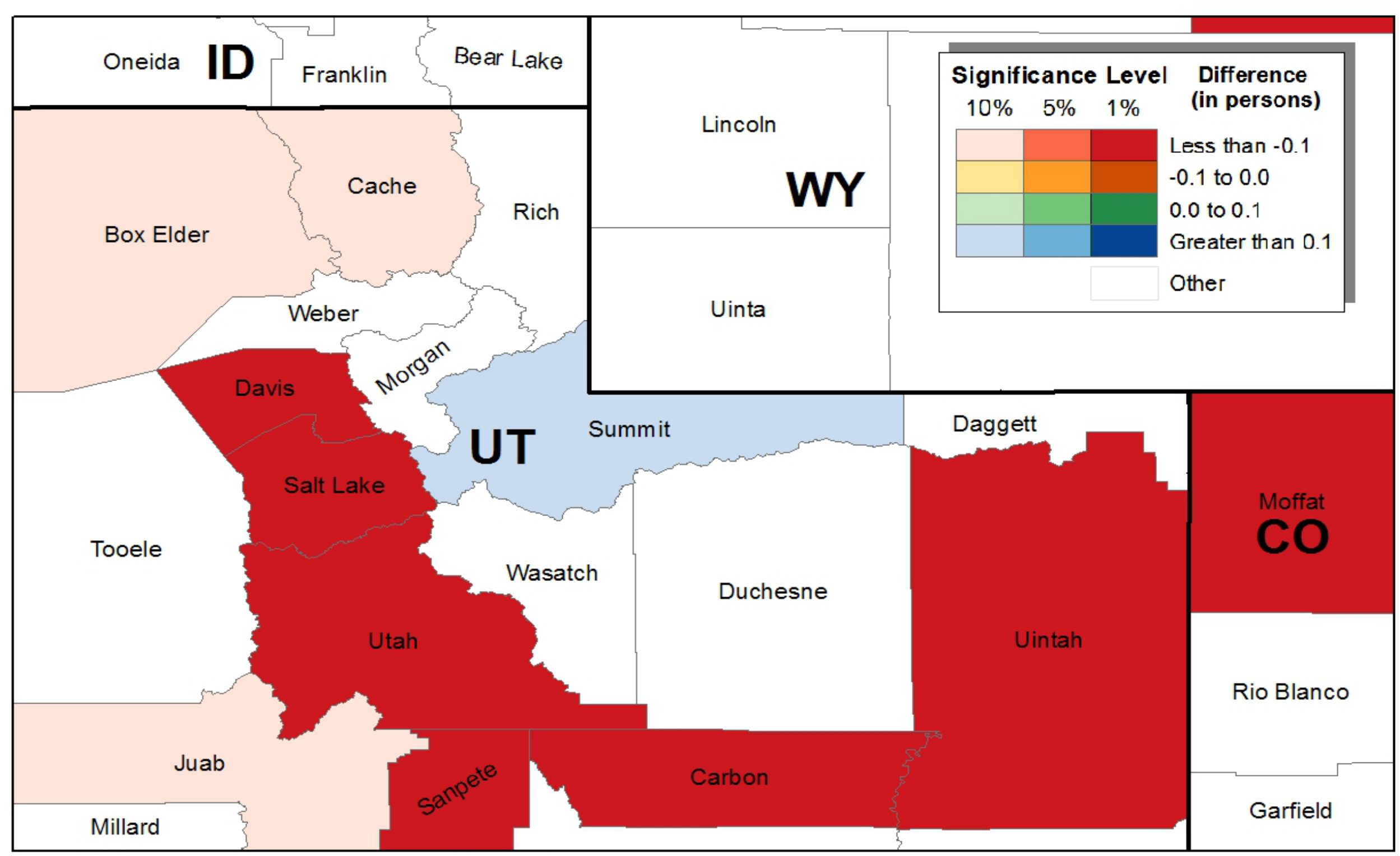

Figure 11. Significant Differences in Vacancy Rates: 2005 ACS Estimates Less Census 2000.

***Significant Differences in Vacancy Rates:***
***2005 ACS Estimates Less Census 2000***

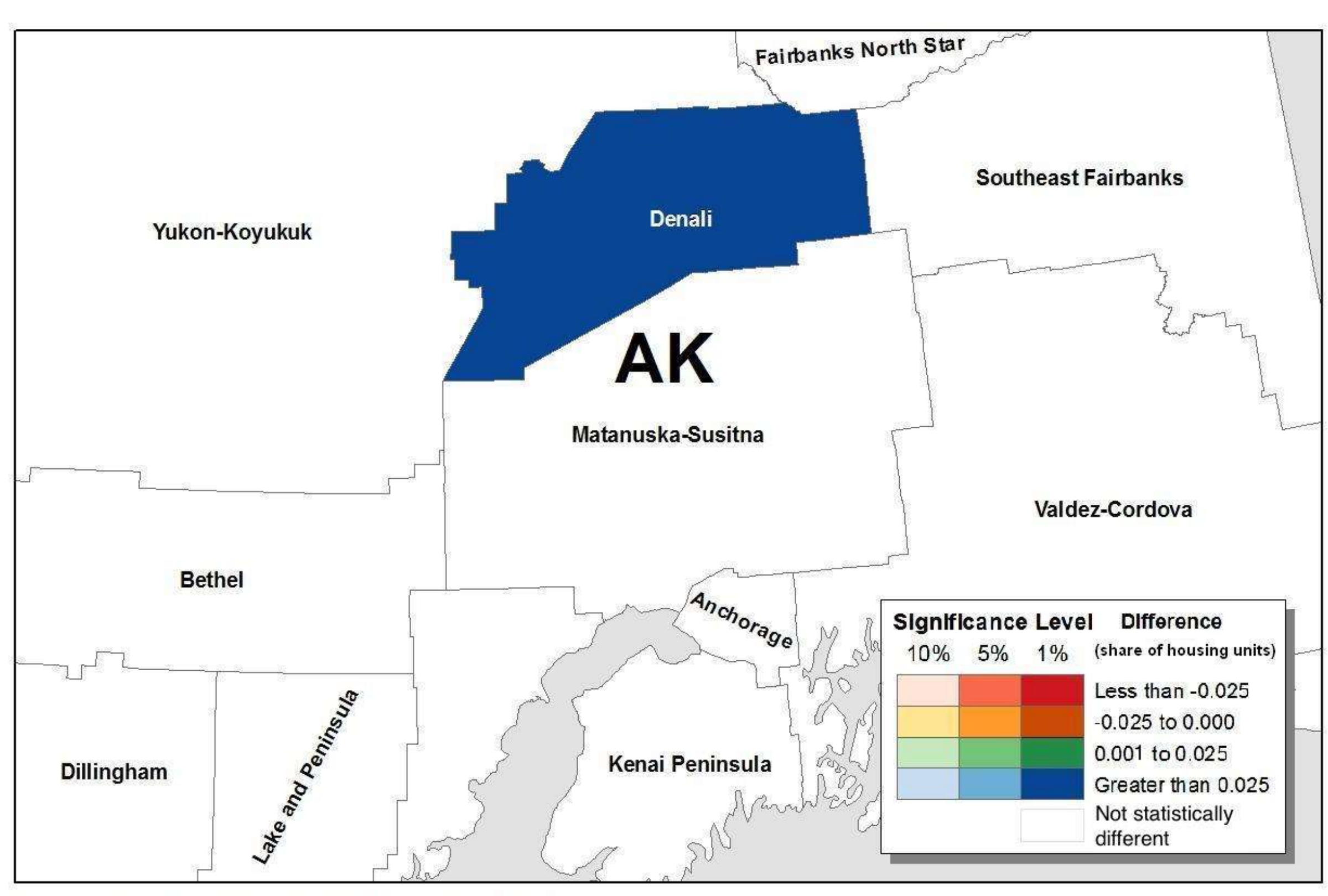